\documentclass[nohyperref]{article}

\usepackage{microtype}
\usepackage{graphicx}
\usepackage{subfigure}
\usepackage{booktabs} 

\usepackage{hyperref}

\usepackage[accepted]{icml2023}

\usepackage{amsmath}
\usepackage{amssymb}
\usepackage{mathtools}
\usepackage{amsthm}

\usepackage[capitalize,noabbrev]{cleveref}

\theoremstyle{plain}

\theoremstyle{definition}

\theoremstyle{remark}

\usepackage[textsize=tiny]{todonotes}

\icmltitlerunning{Co-creating a globally interpretable model with human input}

\begin{document}

\twocolumn[
\icmltitle{Co-creating a globally interpretable model with human input}



\icmlsetsymbol{equal}{*}

\begin{icmlauthorlist}
\icmlauthor{Rahul Nair}{ibm}
\end{icmlauthorlist}

\icmlaffiliation{ibm}{IBM Research Europe, Dublin, Ireland}

\icmlcorrespondingauthor{Rahul Nair}{rahul.nair@ie.ibm.com}
\icmlkeywords{interpretable models; human inputs; decision rules}

\vskip 0.3in
]



\printAffiliationsAndNotice{}  

\begin{abstract}
We consider an aggregated human-AI collaboration aimed at generating a joint interpretable model. The model takes the form of Boolean decision rules, where human input is provided in the form of logical conditions or as partial templates. This focus on the combined construction of a model offers a different perspective on joint decision making. Previous efforts have typically focused on aggregating outcomes rather than decisions logic. We demonstrate the proposed approach through two examples and highlight the usefulness and challenges of the approach.
\end{abstract}
\section{Setting}
\label{sec:intro}

An interpretable model, one that can be understood by humans, offers several desirable attributes. In high risk application domains \cite{eu2021proposal} they can address regulatory and compliance needs, offer safe guards against undesirable outcomes, and augment human understanding of an automated decision process.

We consider building an interpretable model for a supervised classification task jointly with humans. This offers a different perspective in aggregated human-AI decision making, in that the focus is on combining decision logic rather than decision outcomes. We leverage an interpretable model that humans can comprehend, and contribute to, through logical conditions. 

Ensembling decision logic in this way is motivated by several factors. 
(a) \emph{Complementarity:} In generating a prediction/outcome, humans leverage broader context and as such observe a different feature space than those present in the training dataset. Further, humans work with an unknown objective that can be complex. To the extent that the decision process can be codified (evidence suggests this is a hard task \cite{agrawal2019exploring}), this offers potential to improve machine generated logic since it would be unlikely to learn this from data alone. (b) \emph{Interpretabability:} The joint decision process retains the desirable property of interpretability. (c) \emph{Coverage:} Human inputs that capture changes in regulatory requirements, address situations unseen in training data, or limit undesirable behaviours broaden the scope of the aggregated decision process as a whole.

There are several existing notions of hybrid human-AI decision making and how they are achieved technically. A typical setting is one where the machine handles the `easy' cases and defers to human experts for the complicated or uncertain cases \cite{madras2018predict}. Some view human inputs purely as a post-processing step \cite{daly2021user}, where machine outputs are corrected based on feedback. Several works have formalized aggregation. \citep{rastogi2022unifying} provide a unified framework through complementary skills, where the joint decisions are better than those made individually. \citep{wilder2020learning} train a model and delegation decisions jointly. More broadly, there is a growing interest in working logical constraints into ML models \cite{Giunchiglia2022deep}. \cite{zhu2018value} offer a broader lens on working in tacit human knowledge into algorithms through value-sensitive design. 

In this work, we build on previous work on Boolean decision rules \cite{dash2018boolean}. They use a mixed integer program to determine a concise discriminative rule between classes. The decision model takes the form of a IF-THEN-ELSE statement, where the IF condition is in disjunctive normal form (DNF), i.e., clauses that are ORs-of-ANDs. The aim of their method is to determine a concise yet accurate rule set that captures training samples. In our extension, human inputs are stated as rules which are included as soft constraints during the search procedure. The optimal rule set therefore combines evidence from the training data and human inputs.

We briefly note the implications of the choice of using logical relations as an interface between machines and humans. In many cases, for example those stemming from regulatory, safety perspectives, or standard business processes, such a representation can be relatively straightforward to derive. In other cases, however, codifying human judgements as rules can be problematic \cite{agrawal2019exploring}. Moreover, human logic needs to be expressed in terms of the feature space of the training data. This can be a challenge in some cases.

\section{Approach}
\label{sec:model}

\cite{dash2018boolean} formulate the search for an accurate and concise interpretable rule set as a mixed-integer linear programming (MILP) problem. 

To describe the formulation, let $P$ denote the set of positive samples, and $Z$ denote negative samples in the training set $X$. $X$ is assumed to be binarized, i.e., categorical features are one-hot encoded and continuous features are binned. Let $K$ be the set of (exponentially many) possible clauses, namely conjunctions of the binary features in $X$. Define $K_i, K_i\subseteq K$ as the subset of clauses satisfied by observation $i$. 

The main decision variables are $w_k$ for all $k$ in set $K$ --- a binary variable indicating whether conjunction $k$ is selected for the model. Each conjunction $k$ in $K$ has an associated complexity $c_k$. We take $c_k$ to be the degree of the conjunction, i.e.,~the number of participating literals. The formulation also defines $\xi_i$ for $i \in P$ (i.e.,~for all positive samples) to indicate incorrect classification, i.e.,~a false negative.

The objective seeks to minimize total Hamming loss, where the Hamming loss for each sample is the number of conjunctions that must be added or removed to classify it correctly. Specifically, this is expressed as
\begin{equation}
\min_{\xi, w} \sum_{i\in P} \xi_i+ \sum_{i\in Z}\sum_{k\in K_i} w_k.
\label{eq:model:brcg:obj}
\end{equation}
The first term represents the false negatives and the second term the false positives for a choice of conjunctions. False positives add more than `one unit' if they satisfy multiple selected conjunctions, all of which must be removed to classify the instance correctly. This objective is subject to constraints:
\begin{align}
            \xi_i + \sum_{k\in K_i} w_k &\ge 1 \quad \xi_i \ge 0, & \, \forall i\in P \label{eq:model:brcg:c1} \\
             \sum_{k\in K} c_k w_k &\le C \label{eq:model:brcg:c2}\\
             w_k &\in \{0, 1\} & \forall k\in K. \label{eq:model:brcg:c3}
\end{align}
Constraint \eqref{eq:model:brcg:c1} states that for each positive sample, we either have a false negative ($\xi_i = 1$) or include a rule that correctly represents this observation (i.e. a conjunction from the set $K_i$). Constraint \eqref{eq:model:brcg:c2} bounds the total complexity of the selected rule set by a parameter $C$. Constraint \eqref{eq:model:brcg:c3} restricts the decision variables $w_k$ to be binary. 

Problem \eqref{eq:model:brcg:obj}--\eqref{eq:model:brcg:c3} is intractable as written, even with advanced MILP solvers, because the set $K$ is very large and it is prohibitive to generate the entire set of conjunctions. In any case, only a few $w_k$ tend to be selected in the final solution. The authors \cite{dash2018boolean} use a column generation (CG) procedure, which is an iterative algorithm by which candidate conjunctions are generated at each iteration only if they can improve the overall objective. 
The method to generate a new candidate is called the pricing problem and the original model \eqref{eq:model:brcg:obj}--\eqref{eq:model:brcg:c3} is called the master problem. The CG procedure can be summarized by the following steps:
\begin{enumerate}
    \item Restrict the master problem to a small subset of conjunctions $J \subset K$ and solve its linear programming (LP) relaxation obtained by relaxing the constraint $w_k \in \{0,1\}$ to $w_k \geq 0$.
    \item Solve the pricing problem to find conjunctions omitted from $J$ that can improve the objective. Add these conjunctions to $J$.
    \item Repeat steps 1 and 2 until no improving conjunctions can be found.
    \item Solve the unrelaxed master problem ($w_k \in \{0,1\}$) restricted to the final subset $J$.
\end{enumerate}
We refer to \cite{dash2018boolean} for the formulation of the pricing problem and more details in general.

\subsection{Human input}
\label{subsec:human}

Assume some rules for the task are known, either through domain-knowledge, business process rules, regulatory criteria, or safety considerations. As an example, in a mortgage approval task, a condition
\begin{equation*}
    (\text{LTV} \ge 90\%) \lor (\text{LoanAmount} \ge 3.5 \times \text{Income}) 
\end{equation*}
\noindent can be used to reject an application based on Loan-to-Value (LTV) ratio and limits on loan amounts set by regulators. 


Take $U$, $U \subset K$, to be a set of known conjunctions. We modify objective \eqref{eq:model:brcg:obj} as
\begin{equation}
			\min_{\xi, w} 	\color{black}\underbrace{\color{black}\sum_{i\in P} \xi_i + \sum_{i\in Z}\sum_{k\in K_i} w_k}_{\substack{\mbox{Machine objective}\\\mbox{(Hamming loss)}}}  \color{black}+
			\color{black}\underbrace{\color{black}c_u\color{black} n\sum_{k\in U} (1 - w_k).}_{\substack{\mbox{Human inputs}\\\mbox{(violation penalty)}}}
			\color{black}
			\label{eq:user:fulls}
		\end{equation}
This aggregated objective can be interpreted as follows. The model imposes a penalty each time a user-provided input rule is not selected. The latter term serves as regularization and $c_u$ is the lagrange multiplier as a fraction of the dataset size $n$. Since the first two terms in the objective represent the Hamming loss, $c_un$ can be interpreted as the additional Hamming loss that is incurred before a user provided constraint is dropped from the model.

A variant that merits consideration is when human inputs are not known precisely. This occurs when the knowledge of the task is incomplete or rules are only partially known. Denote $U^\prime$ as a set of such partial conditions. This set can also be viewed as a template. We define a distance metric $d(k, U^\prime)$ which computes how similar a conjunction is to provided templates. The objective \eqref{eq:model:brcg:obj} can be rewritten as
\begin{equation}
\min_{\xi, w} \sum_{i\in P} \xi_i + \sum_{i\in Z}\sum_{k\in K_i} w_k + \\
\color{black}\underbrace{\color{black} + c_p\sum_{k\in K}d(k, U^\prime)w_k.}_{\substack{\mbox{distance penalty}\\\mbox{for partials}}}
			\color{black}
\label{eq:user:partials}
\end{equation}
This penalizes conjunctions that are not like those provided in the template set. We note, the parameter $c_p$ does not have the same interpretation as $c_n$ in \eqref{eq:user:fulls}. The distance metric $d(k, U^\prime)$ essentially compares a conjunction to a set of conjunctions. The metric can be computed either by comparing rule semantics or by statistical means if a supporting dataset is available \cite{nair2021changed}.

\subsection{Evaluation}
\label{subsec:model:eval}

The resulting model should generalise well and offer suitable interpretability for domain users. Evaluating a model comes with some subtle challenges in this setting.  For generalisation, typical metrics like test accuracy work. However, one can only draw conclusions if there is supporting data. If the rationale for human inputs is to capture conditions not available in the data, then test data sets may also not reflect such conditions. This occurs frequently in practice, e.g. changing business conditions (increase in LTV values to qualify for a mortgage). Standard generalisation metrics will not offer the full picture in these cases.

Interpretability is generally considered as being the length of the rule set. Shorter rules are preferable and considered more interpretable. One additional measure of interpretability in this setting is related to the content of the rules. If domain-specific inputs are known then one measure is how well the rules mimic domain-specific semantics that are familiar to experts. Such a measure would penalise rule sets using negations for example. We use a semantic rule similarity proposed in \citep{nair2021changed} that solves an assignment problem to determine the least cost mapping between two rule sets. A rule similarity of $1.0$ indicates perfect agreement in semantics and a value $0$ indicates no common semantics. 
\section{Examples}
\label{sec:example}

\subsection{Tic-Tac-Toe}
\begin{figure}[ht]
    \centering
        \begin{tikzpicture}[line cap=round,line join=round]
        \tikzset{every node}=[font=\small\sffamily]
        \tikzset
            {%
              pics/matrix/.style n args={6}{
                code={%
                  \begin{scope}[y=-1cm,scale=0.5]
                    \draw[gray]    (0,0) grid (3,3);
                    \draw[red,line width=1] (#1,#2) -- (#5,#6);
                    \node[blue] at (#1,#2) {x};
                    \node[blue] at (#3,#4) {x};
                    \node[blue] at (#5,#6) {x};
                  \end{scope}     
                }},
            }
          \pic at (0,0)       {matrix={0.5}{0.5}{0.5}{1.5}{0.5}{2.5}};
          \pic at (1.8,0)     {matrix={1.5}{0.5}{1.5}{1.5}{1.5}{2.5}};
          \pic at (3.6,0)     {matrix={2.5}{0.5}{2.5}{1.5}{2.5}{2.5}};
          \pic at (5.4,0)     {matrix={0.5}{2.5}{1.5}{1.5}{2.5}{0.5}};     
          \pic at (0,1.8)     {matrix={0.5}{0.5}{1.5}{0.5}{2.5}{0.5}};
           \pic at(1.8,1.8)   {matrix={0.5}{1.5}{1.5}{1.5}{2.5}{1.5}};
          \pic at(3.6,1.8)    {matrix={0.5}{2.5}{1.5}{2.5}{2.5}{2.5}};
          \pic at(5.4,1.8)    {matrix={0.5}{0.5}{1.5}{1.5}{2.5}{2.5}};
              
        \end{tikzpicture}
    \caption{The 8 deterministic rules that classify `x' as winning a tic-tac-toe game.}
    \label{fig:tictactoe}
\end{figure}
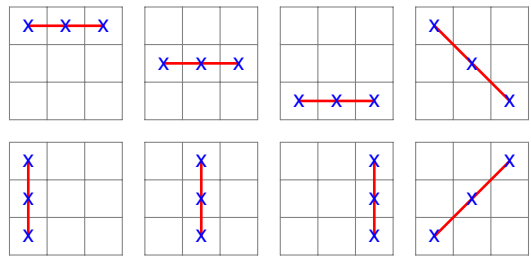

The key ideas are first shown on the game of tic-tac-toe. Given the state of the end board\footnote{https://archive.ics.uci.edu/ml/datasets/Tic-Tac-Toe+Endgame}, consider the task of predicting if `x' wins the game. In tic-tac-toe, there are exactly eight rules under which a model predicts \texttt{true}: three `x' verticals, three horizontals, and two diagonals (Figure \ref{fig:tictactoe}). Irrespective of the state of the rest of the board, if any of these eight conditions occur, the prediction must be true. This is a noise-free classification task with deterministic rules.

We aim to learn these rules from data on the end board configuration under varying levels of human inputs.  While the rules are picked randomly, they are assumed to be known perfectly. As \cite{dash2018boolean} report, their algorithm is able to recover all rules from this dataset. We therefore limit the data available to the algorithm as well, to degrade machine performance for illustrative purposes.

\begin{figure}[ht]
    \centering
    \includegraphics[width=\columnwidth]{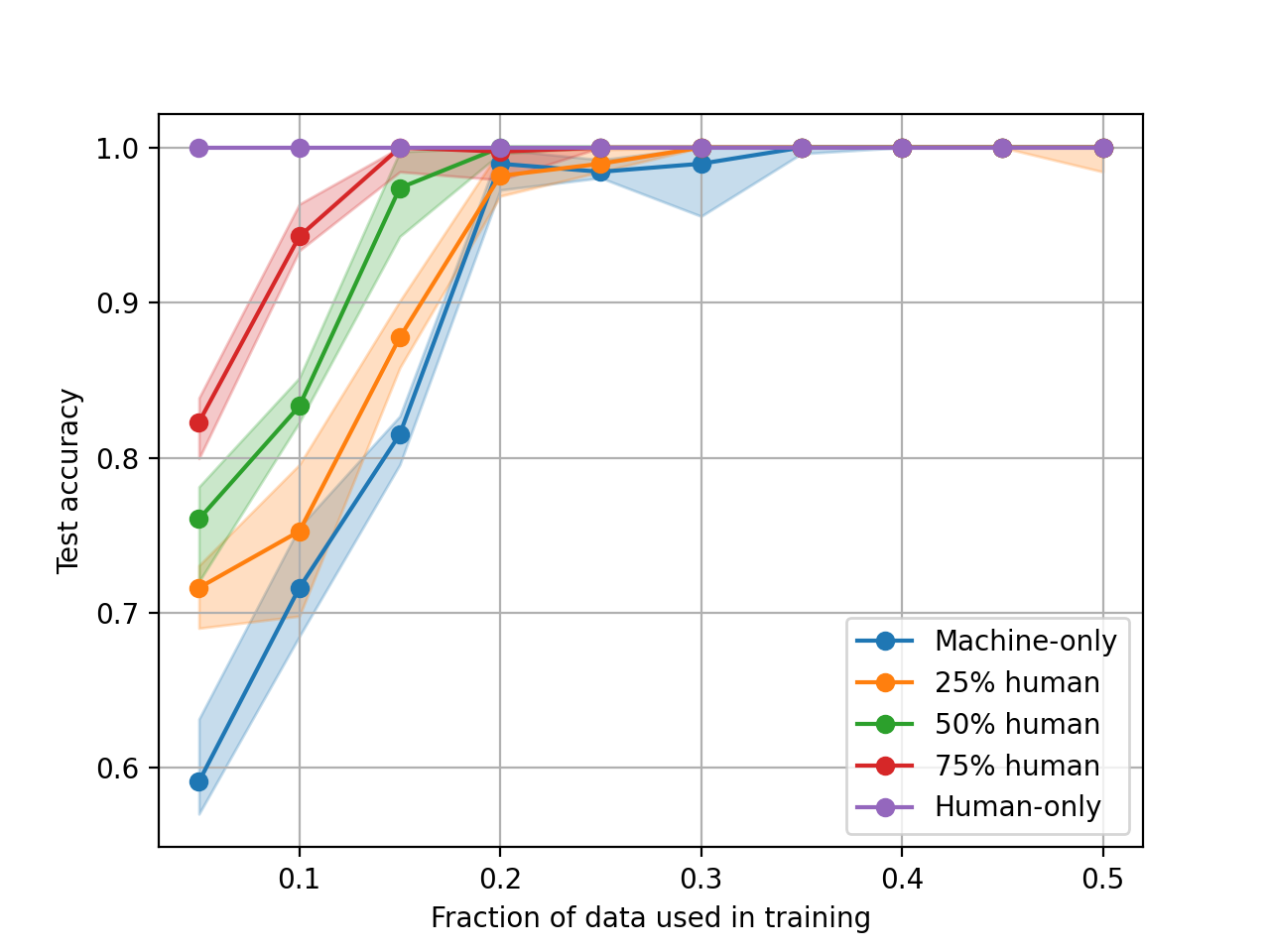}
    \caption{Median test accuracy (with 25th and 75th percentiles) from a 5-fold cross-validation, shown as a function of training size and size of human inputs for tic-tac-toe end game.}
    \label{fig:ex:tictactoe-acc}
\end{figure}

\begin{figure}[ht]
    \centering
    \includegraphics[width=\columnwidth]{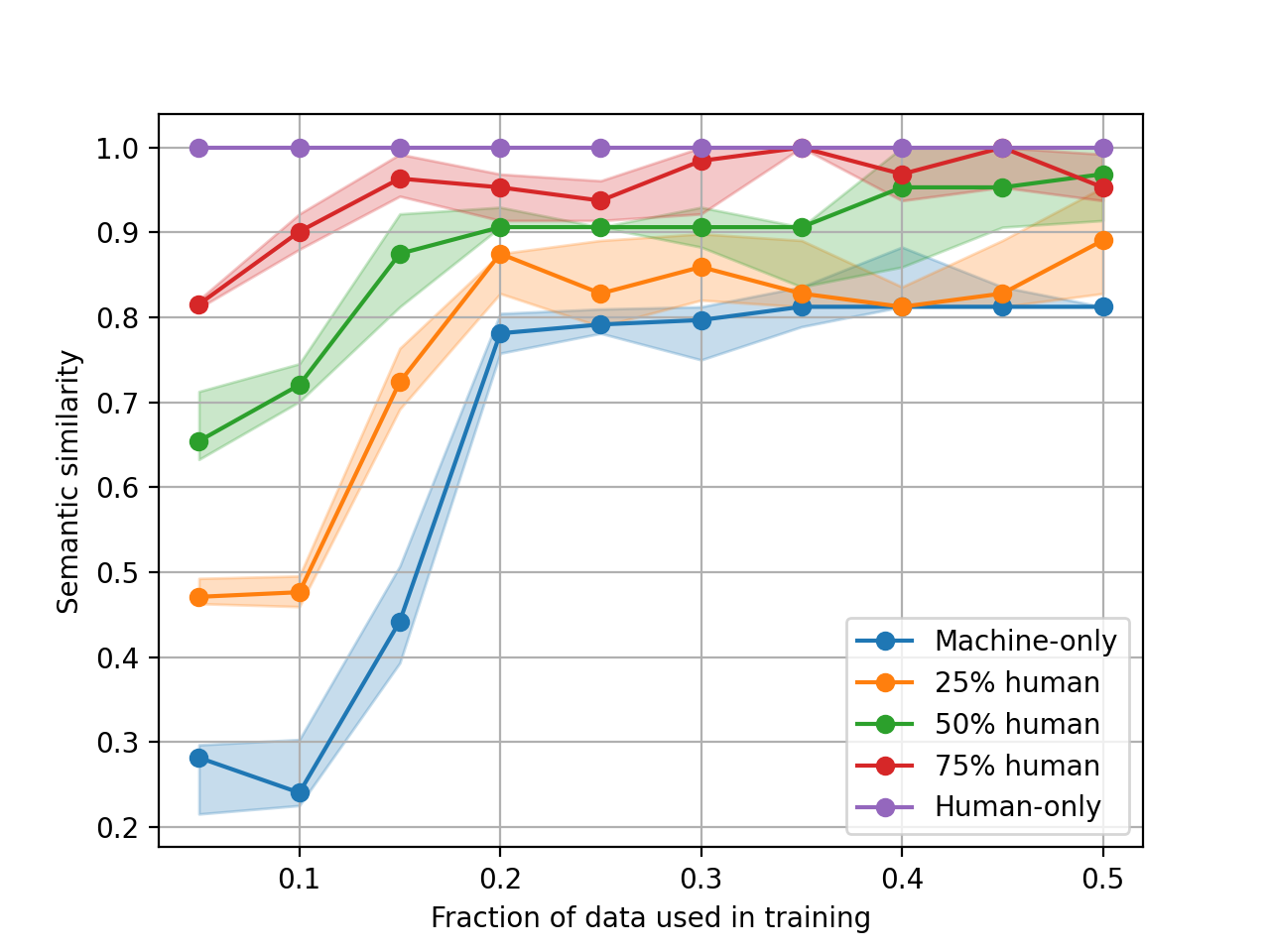}
    \caption{Semantic rule similarity (with 25th and 75th percentiles) from a 5-fold CV. Higher is better.}
    \label{fig:ex:tictactoe-sim}
\end{figure}

We implement the model described by objective \eqref{eq:user:fulls} and constraints \eqref{eq:model:brcg:c1}--\eqref{eq:model:brcg:c3} and use CPLEX as the solver. We measure the test accuracy (Figure \ref{fig:ex:tictactoe-acc}) and semantic rule similarity (Figure \ref{fig:ex:tictactoe-sim}) on a hold out set over five folds. The complexity parameter $C$ is set to $24$, and $c_n=0.05$. We vary the amount of training data available to the method along with extent of human input, i.e. either no inputs, all 8 rules or partial rules of 2, 4 or 6 randomly sampled rules.

The effects are most pronounced when the algorithm has access to only 5\% of the data. Here, an machine-only solution has a median accuracy of 59\%.  The algorithm generates a perfect model for the case when the human inputs all 8 rules with perfect information about the game. When the rules are partial, results fall in between these two extremes. 

Rule semantics present a similar story shown in Figure \ref{fig:ex:tictactoe-sim}. Machine-only generated rule sets exhibit poor semantic similarity to the 8 known win rules and improve with additional human-provided rule sets. However, even in this simple case, without significant human-guidance, learnt rules are expressed differently even if they have the same implication. For instance, the diagonal rules are often expressed as the absence of an `o' rather than a presence of an `x'. This illustrates that the human-AI model outperforms the machine-only outcomes in generalisation and rule semantics.


\subsection{Healthcare}
\label{subsec:apps}

We now consider a more practical case in healthcare. Sepsis is a serious medical condition due to organ dysfunction brought about by infection \cite{singer2016third}. It is associated with fast progression of the condition and a high risk of death. Rapid treatments are a key way to better manage the condition and reduce mortality risk. Several aspects of sepsis are not very well understood \cite{hotchkiss2003pathophysiology}.

In practice, scoring systems like SIRS and qSOFA are used to assess if sepsis likely so treatment can be initiated quickly. These methods use logical conjunctions based a patient's vital signs. qSOFA suspects sepsis if two of the three conditions on respiratory rate ($\ge 22$ breaths per minute), blood pressure ($\le 100$ mmHg), or altered mental state are true.

Treatments vary by severity of sepsis which is generally classed in three categories (sepsis, severe sepsis, and septic shock). Several drug regimes exist to treat sepsis. While anti-bacterials target source of the infection, mild and severe sepsis are each treated with different medications. Drugs meant for severe sepsis cannot be administered before those used for mild sepsis, as this can lead to antibiotic resistance \cite{li2020temporal}. Another class of drugs (vasoactive and diuretics) are used to reduce severe symptoms. These considerations form the basis of expert human input. 

We consider the task of predicting survival condition of patients using 30-day mortality outcomes. From the MIMIC-III (v1.4) dataset \cite{johnson2016mimic}, we extract data on ICU admissions with suspected sepsis (1,783 admissions). For these admissions we retrieve age and gender of the patients along with several vitals such as respiration, temperature, etc. We include medications relevant to sepsis that were administered over the course of their ICU stay. Similar to \cite{li2020temporal}, we consider $3$ drugs for severe sepsis, $7$ for mild sepsis, $3$ vasoactive, and $1$ diuretic medication. 

Our setting differs from that of \citep{li2020temporal} in that the temporal dynamics are simplified in the following manner. Abnormalities in vital signs are aggregated over a day. Missing vital signs are considered to be normal. The survival condition is measured at 30 days after hospital admission. 

\paragraph{Human input} For our experiments, we consider the conditions on administering drugs in a specific order similar to \cite{li2020temporal}, i.e.
\begin{equation*}
    \text{Survival Outcome} \leftarrow \text{Use Drug}_1 \land \text{Use Drug}_2
\end{equation*}
where, Drug$_1$ denotes set of drugs used for severe sepsis, and Drug$_2$ are drugs for mild sepsis. 

\begin{figure}[t]
    \centering
    \includegraphics[width=\columnwidth]{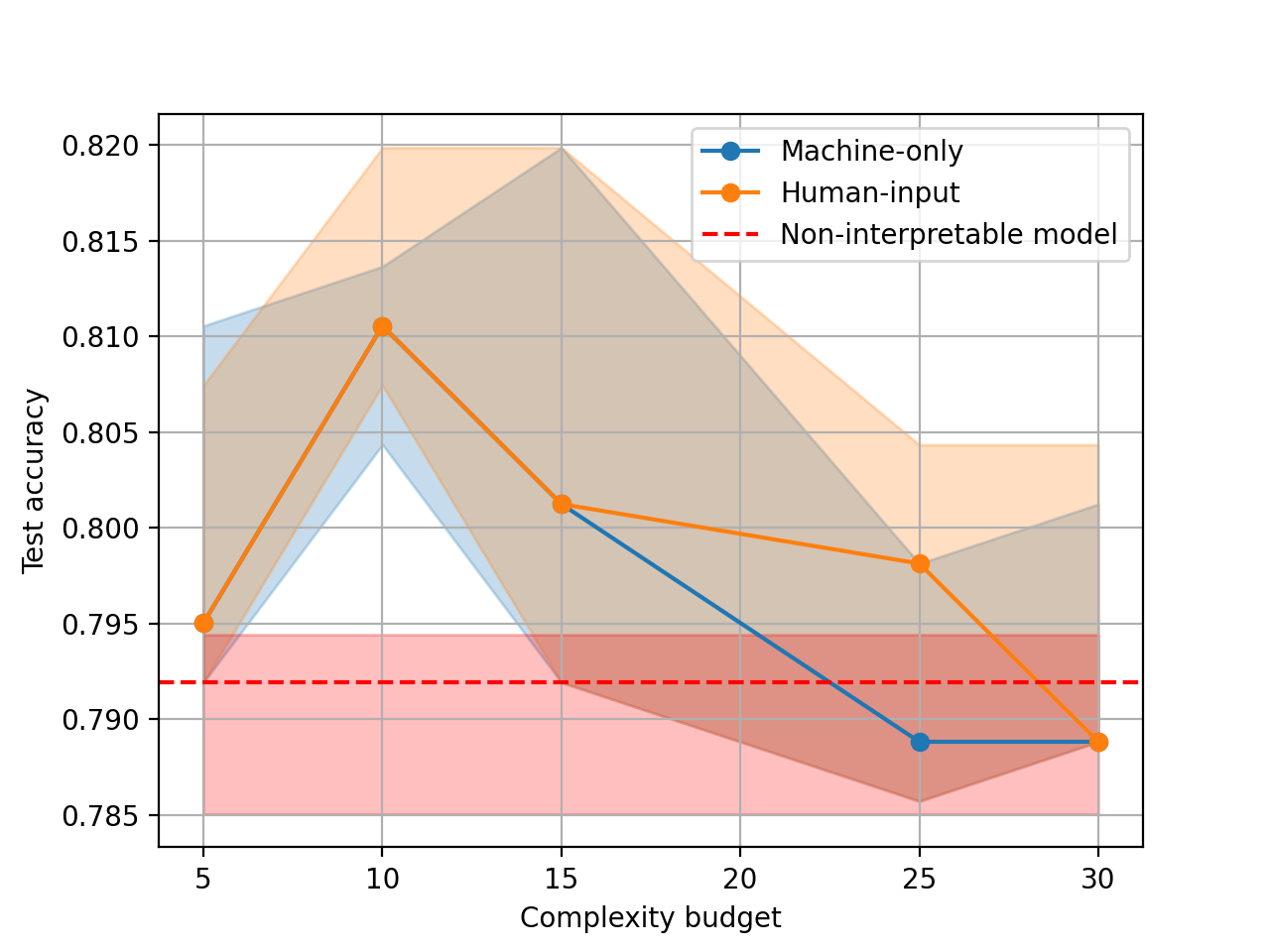}
    \caption{Median test accuracy (with 25th and 75th percentiles) from a 5-fold cross-validation}
    \label{fig:ex:sepsis}
\end{figure}

\begin{figure}[ht]
    \centering
    \includegraphics[width=\columnwidth]{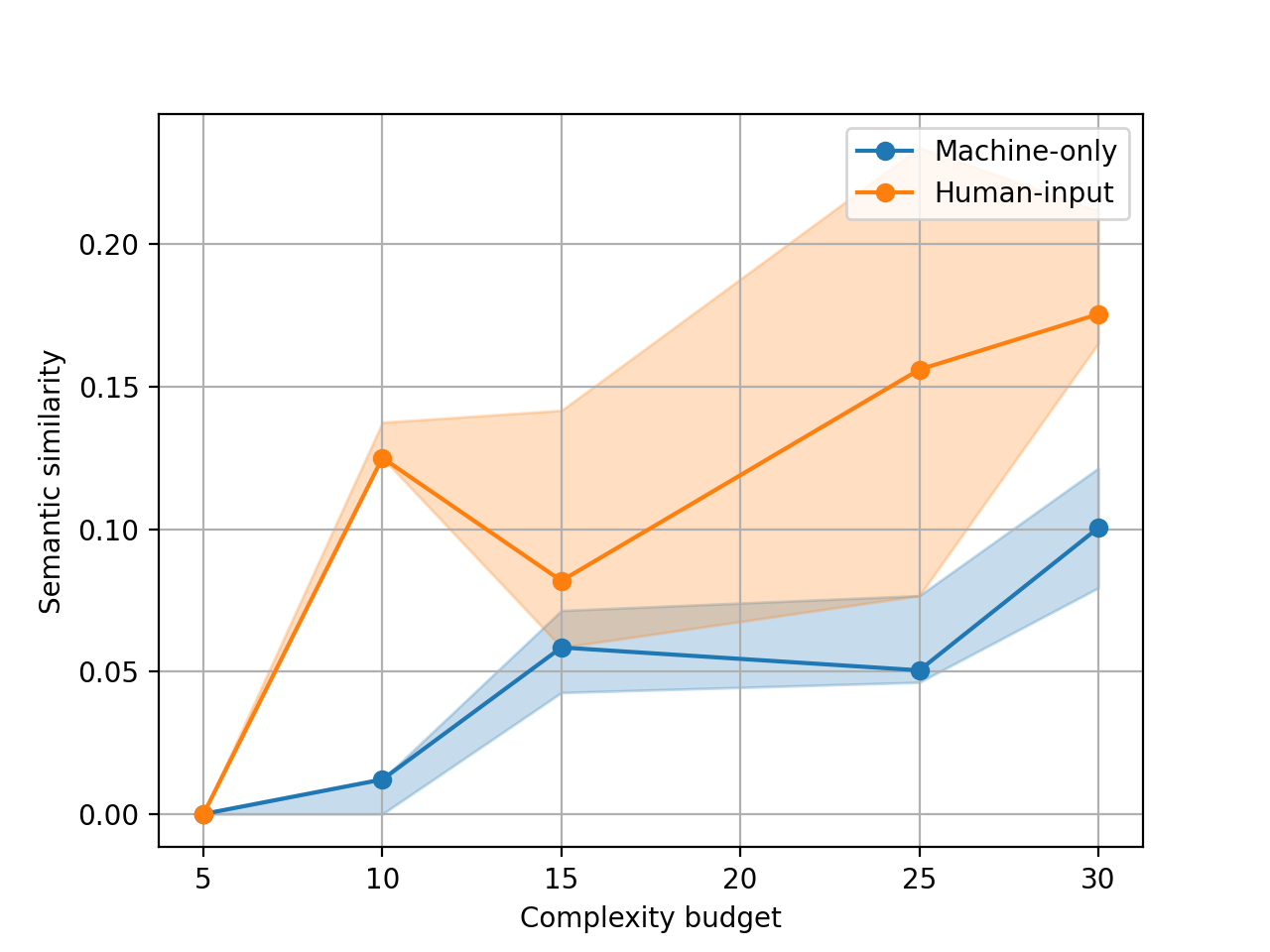}
    \caption{Semantic rule similarity (with 25th and 75th percentiles), 5-fold CV showing }
    \label{fig:ex:sepsis-sim}
\end{figure}

Three models were trained. One `machine-only' model, a second rule induction with human input, and a third non-interpretable model as a baseline. The first two were run for several values of the complexity budget $C=\{5, 10, 15, 25, 30\}$. The non-interpretable model is a Random Forest from scikit-learn with all default values \cite{scikit-learn}. All models were run over 5 cross-validation folds. The two metrics of interest, test accuracy (Figure \ref{fig:ex:sepsis}) and rule similarity (Figure \ref{fig:ex:sepsis-sim}) computed. Rule similarity here is relative to human-input. 

The performance of both interpretable models is better than the baseline RF model for most cases. The human-assisted variant does no worse than the machine-only variant. When $C=25$ it offers some gains. Despite the performance being similar, rule similarity relative to human-input is better. What does this mean? On one hand it is somewhat expected, as human expert rules were provided as input to one and not the other. However, considering similar performance, this suggests that the human-input variant used domain-specific logic instead of statistical basis to construct its rule set. The main advantage therefore is that the resulting model reflects domain-expert intuition and insight. 

A sample rule set for the best performing model when $C=10$ which achieves 82.9\% accuracy is shown below.
\begin{equation*}
\begin{split}
    \text{Positive } & \text{Outcome} \leftarrow \\
    &(\text{Use Vancomycin} \land \text{Use Metronidazole}) \lor \\
    &(\neg(\text{severe+mild}) \land \neg (\text{Use Furosemide}) \land \\
    & \text{Normal CRP} \land \text{Abnormal Creatinine} \land \\
    & \text{Use Norepinephrine} \land \text{Abnormal Lactate} \land \\
    & \neg(\text{Use Metoprolol}) \land (\text{Age}>52.0) )
\end{split}
\end{equation*}

The rule set consists of two conjunctions. The first is from human-inputs and refers to the drug combination for patients with severe and mild sepsis. The second refers to a cohort with the same severity throughout their admission and specific vitals, e.g. abnormal lactate, and with specific medications. We caution that this rule set is illustrative only and without clinical validation.


\section{Discussion}
\label{sec:discussion}

A model to generate a joint interpretable model with human input has been proposed and demonstrated on two examples. The main idea explored is the combination of decision logic directly, as opposed to aggregating outcomes. Aggregating decision logic in this way offers pipelines that reflect expert insights. By construction, the rule sets have bounded complexity and is fully interpretable. This makes it useful for practical applications.

\noindent \emph{Disagreement:} When human inputs directly contradict the data, the optimization model may prefer one or the other depending on the parameter $c_u$. Larger values will make human provided rules more `sticky' in the final model. Preliminary results suggest this is the case; however, it is worth considering other ways to express such cases within the model. 

\noindent \emph{Complementarity:} In some cases, it may be feasible to assess how complementary human and machine inputs are, as done in \citep{rastogi2022unifying}. The main constraint is the availability of evidence to make such an assessment. If it is available, then the manner in which the complexity budget $C$ is shared between human and machine generated conjunctions can provide additional insights. 

\noindent \emph{Satisfiability:} We have explored the case where human inputs are treated as soft constraints, i.e., can be violated. In cases where such inputs need to be satisfied, these can be treated as hard constraints, by including constraints in the optimization model directly. This strictly enforces human inputs.

Additionally, we have touched upon some challenges associated with such an approach. Chief among them is the codifying of human judgements in terms of available attributes. Empirical challenges around evaluations can also be challenging if human inputs are based on exogenous factors not reflected in training data. The use of logical conjunctions in the DNF form uses axis-aligned primitives. This is a natural and flexible approach but has limits when more advanced feature transformations need to be expressed.

\section{Acknowledgements}
Thanks to Jonathan Epperlein 
and 
Anne-Marie Cromack 
for comments and review of an earlier draft of the paper. This work was partially funded 
by the European Union’s Horizon Europe research and innovation programme under grant agreement no. 101070568.

\bibliography{icml2023/biblio}
\bibliographystyle{icml2023}

\end{document}